# Ground State thermodynamic and response properties of electron gas in a strong magnetic and electric field: Exact analytical solutions for a conventional semiconductor and for Graphene

Georgios Konstantinou[1] & Konstantinos Moulopoulos[2]

*Department of Physics, University of Cyprus, PO Box 20537, 1678 Nicosia, Cyprus*

[1]*ph06kg1@ucy.ac.cy*, [2]*cos@ucy.ac.cy*

Consequences of an exceedingly strong electric field (*E* field) on the ground state energetics and transport properties of a 2D spinless electron gas in a perpendicular magnetic field (a Quantum Hall Effect (QHE) configuration) are investigated to all orders in the fields. For a conventional semiconductor, we find fractional values of the Hall conductivity and some magnetoelectric coefficients for certain values of *E* and *B* fields that do not result from interactions or impurities, but are a pure consequence of a strong enough in-plane *E* field. We also determine analytically the ground state energy, and response properties such as magnetization and polarization as functions of the electromagnetic field in the strong *E* field limit. In the case of Graphene, we obtain more complex behaviors leading to the possibility of irrational Hall values. The results are also qualitatively discussed in connection to various mechanisms for the QHE-breakdown.

## 1. Conventional system

### 1.1 Low E-field strength

Consider an ideal nonrelativistic two-dimensional spinless electron gas in a perpendicular and homogeneous magnetic field $\vec{B}$ directed along the positive *z*-axis. The dimensions of the plane are taken to be macroscopically large $L_x \times L_y$. In addition to $\vec{B}$ there is also an in-plane, homogeneous electric field $\vec{E}$ pointing in the *y*-direction that is very weak, i.e. it doesn't cause any overlap of different Landau Levels (L.Ls). In this manner, the Fermi energy can always be situated in the interior of an energy gap (i.e. for certain areal density), causing, as is well-known, the universality of quantization of Hall conductance $\sigma_H$. If the electric field is further increased from its first critical value (to be determined below) then the gap closes and the system is highly nonlinear, the general consensus being that this results to destruction of the quantization of $\sigma_H$, although from what we will see below several refined behavioral patterns remain (in this strong electric field case), that can even lead to the survival or even a different type of integer quantization in the nonlinear regime. This strong electric field case will be separately studied in the next section.

We choose to work in the Landau gauge $A_x = -By$, $A_y = A_z = 0$ along with a scalar potential $V = eEy$ (we take the charge of the electron to be $-e$) resulting in the following Hamiltonian:

$$H = \frac{1}{2m}\left(\vec{p} + \frac{e}{c}\vec{A}\right)^2 + eEy,$$

with energy spectrum [1] $\varepsilon = \hbar\omega(n + \frac{1}{2}) + mV_D^2/2 + eEY_0$ **(1.1)**, with $\omega = eB/mc$ being the cyclotron frequency, $n = 0, 1, 2...$ the L.L. index, $V_D = c\vec{E}\times\vec{B}/B^2 = cE/B$ the modulus of the drift velocity and $Y_0 = \frac{cp_x}{eB} - \frac{mc^2E}{eB^2}$ **(1.2)** the guiding center operator eigenvalue in the $y$-direction (restricted to the area $-L_y/2 \leq Y_0 \leq L_y/2$). Now, in specifying the electric field's strength $E$, if we first want to avoid any overlap among different L.Ls, the single particle energy $\varepsilon(n-1, Y_0 = L_y/2)$ must be lower than (or equal to) $\varepsilon(n, Y_0 = -L_y/2)$, a criterion which leads to the following inequality: $E \leq \hbar\omega/eL_y$ **(1.3)**. Therefore, this case of no-overlap involves a limitation of drift velocity's values that depends on both field strengths and is equivalent to either of two criteria: (i) $mV_D L_y \leq \hbar$ (the angular momentum of an electron in one edge with respect to a point in the other edge is $\leq \hbar$), or (ii) $mV_D^2/2 \leq \hbar^2/2mL_y^2$ (the drift kinetic energy is $\leq$ a confinement energy along the $y$-direction (due to the uncertainty principle)).

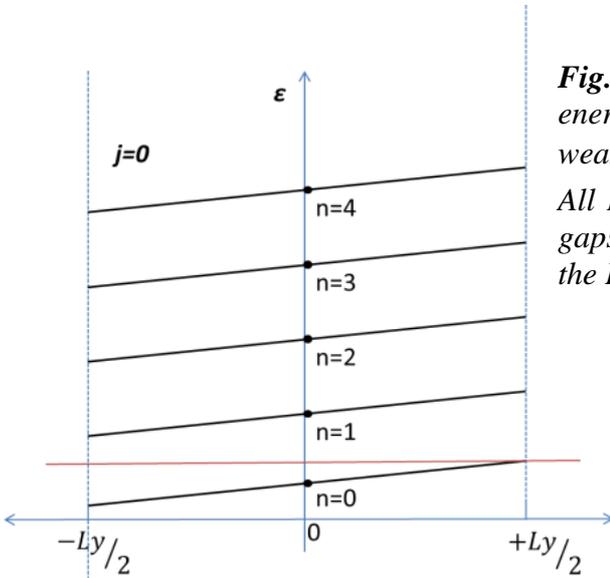

*Fig. 1.1: A schematic representation of the energy states (L.Ls) when the electric field is weak $E \leq \hbar\omega/eL_y$: there is no inter-L.L. overlap. All L.Ls can be filled independently, and energy gaps between L.Ls can result in quantization of the Hall conductivity.*

In this case all L.Ls can be filled independently: according to the least energy principle, at zero temperature $T=0$, all electrons occupy states labeled by small quantum numbers $(n, l)$, (with $n$ the above mentioned Landau Level index, a non-zero positive integer, and with $l$ another integer specifying the eigenvalue of $p_x$ due to periodic boundary conditions along the *x*-direction, the eigenvalue being $p_x = hl/L_x$), starting from L.L $n=0$ and varying all possible values of *l*, and then successively following *n=1*, *n=2* and so on. Each L.L (indexed by quantum number *n*) may host up to $\Phi/\Phi_0$ spinless electrons (which is the total number of distinct values of *l* in the thermodynamic limit) with $\Phi = BL_xL_y$ being the magnetic flux across the 2D plane and $\Phi_0 = hc/e$ the flux quantum, in order to be consistent with Pauli principle. This means that without loss of generality we always have, let us say, *ρ* (=1,2,3..) L.Ls that are occupied for fixed particle number *N* and for a certain value of *B,* and due to the minimal energy criterion, we must have (*ρ-1*) fully occupied L.Ls and the last L.L. either partially or fully occupied. Of course, if *N* is fixed, the entire occupational procedure is uniquely determined only by the magnetic field strength **B** and it can be described naturally by the inequalities

$$(\rho - 1)\frac{\Phi}{\Phi_0} \leq N \leq \rho \frac{\Phi}{\Phi_0} \Rightarrow \frac{1}{\rho} n_A \Phi_0 \leq B \leq \frac{1}{\rho - 1} n_A \Phi_0, \quad (1.4)$$

with $n_A = N/L_xL_y$ the 2D areal density of electrons. The last L.L. is occupied by $N - (\rho - 1)\Phi/\Phi_0$ electrons allowing us to determine analytically the maximum $Y_{0MAX}$ of the last electron located at L.L. index value of *n=ρ-1*:

$$Y_{0MAX} = \frac{chN}{eBL_x} + L_y\left(\frac{1}{2} - \rho\right) \quad (1.5)$$

which gives the expected results that when $N = \rho\Phi/\Phi_0$, $Y_{0MAX} = L_y/2$ and when $N = (\rho - 1)\Phi/\Phi_0$, $Y_{0MAX} = -L_y/2$.

## *1.2 Thermodynamic properties*

The total internal energy of the system at *T=0* is a sum over all occupied quantum numbers *n* and $Y_0$:

$$\mathbf{E} = \sum_n \sum_{Y_0} \left[\hbar\omega\left(n + \tfrac{1}{2}\right) + \frac{1}{2}mV_D^2 + eEY_0\right] \quad (1.6)$$

In the macroscopic (continuum) limit $L_y \to \infty$ we may approximate the sum with respect to $Y_0$ with an integral:

$$\sum_{Y_0 = -L_y/2}^{L_y/2} f(Y_0) \to \frac{BL_x}{\Phi_0} \int_{-L_y/2}^{L_y/2} dY_0 f(Y_0), \text{ with } f(Y_0) \text{ an arbitrary function of } Y_0. \quad (1.7)$$

We then have

$$\mathbf{E} = \frac{BL_x}{\Phi_0} \sum_{n=0}^{\rho-2} \int_{-L_y/2}^{L_y/2} dY_0 \left[\hbar\omega(n+\tfrac{1}{2}) + \tfrac{1}{2}mV_D^2 + eEY_0\right] + \frac{BL_x}{\Phi_0} \int_{-L_y/2}^{Y_{0MAX}} dY_0 \left[\hbar\omega(\rho-\tfrac{1}{2}) + \tfrac{1}{2}mV_D^2 + eEY_0\right]$$

and after a number of algebraic manipulations we reach a closed analytical expression for the total minimal energy per electron, namely

$$\frac{\mathbf{E}}{N} = -\frac{e^2 B^2}{4\pi mc^2 n_A}\rho(\rho-1) + eEL_y\left[\frac{n_A\Phi_0}{2B} - \left(\rho - \frac{1}{2}\right) + \frac{\rho B}{2n_A\Phi_0}(\rho-1)\right] + \frac{mc^2 E^2}{2B^2} + \frac{\hbar eB}{mc}\left(\rho - \frac{1}{2}\right)$$

**(1.8)**

valid in the magnetic field range $n_A\Phi_0/\rho \leq B \leq n_A\Phi_0/\rho-1$. Or, in units of the Fermi energy $\varepsilon_F = \hbar^2 4\pi n_A/2m$ (the one defined for 2D spinless electron gas in the absence of $E$ and $B$),

$$\frac{\mathbf{E}}{N\varepsilon_F} = -\frac{B^2\rho(\rho-1)}{2\Phi_0^2 n_A^2} + \frac{B}{n_A\Phi_0}\left(\rho - \frac{1}{2}\right) + \frac{L_y e n_A \Phi_0 E}{2B\varepsilon_F} - eEL_y\left(\rho - \frac{1}{2}\right) \quad (1.9)$$
$$+ \frac{e^2 E^2}{4\pi\varepsilon_F^2 n_A}\frac{\Phi_0^2 n_A^2}{B^2} + \frac{eEL_y}{2\varepsilon_F}\frac{B}{n_A\Phi_0}(\rho-1)\rho$$

Now, let $B$ be equal to $B = \frac{1}{k}n_A\Phi_0$, with $\rho-1 \leq k \leq \rho$, so that from (1.3) we have $\frac{eEL_y}{\varepsilon_F} \equiv y \leq \frac{1}{k}$. Rewriting then (1.8) in terms of $y$ and $x = B/n_A\Phi_0$ we obtain

$$\frac{\mathbf{E}}{N\varepsilon_F} = -\frac{\rho(1-\rho)x}{2} + x\left(\rho - \frac{1}{2}\right) + \frac{y}{2x} + \frac{1}{4\pi n_A L_y^2}\frac{y^2}{x^2} - y\left(\rho - \frac{1}{2}\right) + \frac{yx}{2}(\rho-1)\rho$$

In Fig. 1.2 see the graphs of energy and magnetization per electron, for $n_A L_y^2 = 1$ (a good value so that the internal structure of these quantities are shown in sufficient detail).

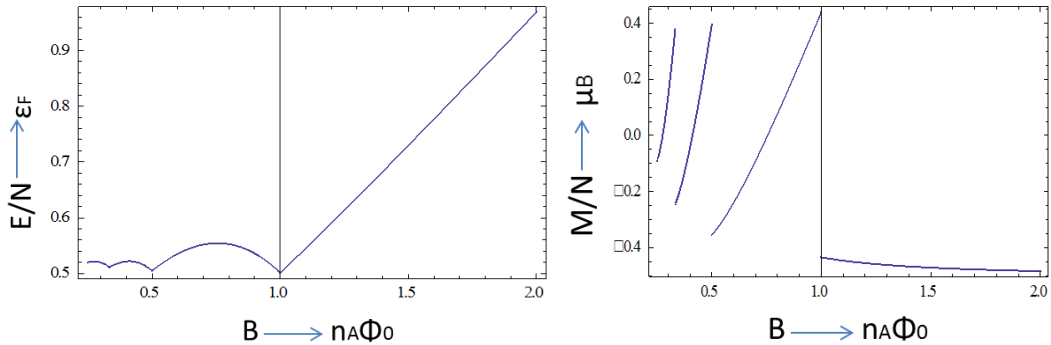

***Fig. 1.2:*** *Energy and Magnetization per electron as functions of the magnetic field B.*

Using the second thermodynamic law we can also determine analytically the equilibrium magnetization and polarization per electron, which turn out to be

$$\frac{\mathbf{M}}{N} = -\frac{\partial(\mathbf{E}/N)}{\partial B} = \frac{e^2 B}{2\pi mc^2 n_A}\rho(\rho-1) - eEL_y\left[-\frac{n_A\Phi_0}{2B^2} + \frac{\rho(\rho-1)}{2n_A\Phi_0}\right] + \frac{mc^2 E^2}{B^3} - \frac{\hbar e}{mc}\left(\rho - \frac{1}{2}\right)$$

$$\frac{\mathbf{P}}{N} = -\frac{\partial(\mathbf{E}/N)}{\partial E} = -eL_y\left[\frac{n_A\Phi_0}{2B} - \left(\rho - \frac{1}{2}\right)\right] - \frac{mc^2 E}{B^2} - \frac{L_y(\rho-1)}{n_A c}\left(\frac{\rho e^2}{2h}\right)B \quad (1.10)$$

These thermodynamic expressions demonstrate the effect of the electric field which is non-linear (with respect to the variable *E*), even for the case of relatively weak electric field. Now we proceed by examining some limits**:** when all L.Ls are fully occupied, namely when $N = \rho\Phi/\Phi_0$, then we obtain

$$\frac{\mathbf{M}}{N} = -\frac{e\hbar}{2mc} + \frac{L_y}{n_A c}\left(\frac{\sigma_H}{2}\right)E + \frac{m}{n_A^3 c}\sigma_H^3 E^2 \quad (1.11)$$

$$\frac{\mathbf{P}}{N} = -\frac{m}{n_A^2}\sigma_H^2 E \quad (1.12)$$

with $\sigma_H = \frac{\rho e^2}{h}$ the Hall conductivity (see below). It is interesting to note a characteristic half of the Hall conductance in the coefficient that connects the so-called magnetoelectric effects with the fields that cause them (that in a problem with chiral properties, usually in Topological Insulator materials, correspond to an extra magnetization caused by a parallel electric field (as in (1.11) above) and an extra polarization caused by an extra magnetic field (see [2] and references therein)**;** here we note such a trend even in a Quantum Hall system (which is not unexpected, and is actually justified based on general Physics arguments [2]). One can actually see such a trend in the polarization as well, for a general *B*, in (1.10), where in the last linear term we can again see σ$_H$/2 appearing. [It should be added that all these behaviors originate from the final term of (1.9) that describes an *EB*–coupling, something that we will also see later for the relativistic case.] In addition to all this, note also that for $E = \hbar\omega/eL_y$ the gap closes, and (1.8) becomes

$$\frac{\mathbf{E}}{N} = \frac{1}{2}\varepsilon_F(E=0, B=0) + \frac{\hbar^2}{2mL_y^2}, \quad (1.13)$$

with $\varepsilon_F(E=0, B=0) = \hbar^2 4\pi n_A / 2m$ being the expected Fermi energy (in the absence of fields) – and when all L.Ls are fully occupied, with $B = n_A\Phi_0/\rho$, and with any *E* (now less than its first critical value) (1.8) becomes

$$\frac{\mathbf{E}}{N} = \frac{1}{2}\varepsilon_F(E=0, B=0) + \frac{mc^2 E^2}{2B^2}$$

i.e. the total energy per electron reduces to 2D energy of free electron gas plus a drift kinetic term. These last simple results are not quite unexpected and may be justified with proper semiclassical considerations.

## *1.3 Hall Conductivity*

The Hall conductivity is defined as $\sigma_H = e n_A c / B$. The usual plot representing the QHE [3] shows $\sigma_H$ as a function of *B*, and in this we have plateau formation for certain values of *B*. For *B* varying in the range defined by the inequaltities (1.4)

$$\frac{1}{\rho} n_A \Phi_0 \leq B \leq \frac{1}{\rho-1} n_A \Phi_0, \quad (1.14)$$

then $\sigma_H$ is quantized in units of $\rho e^2 / h$ with $\rho$ being an integer, which counts the number of fully occupied L.Ls. Note that each plateau has maximum width

$$\frac{1}{\rho-1} n_A \Phi_0 - \frac{1}{\rho} n_A \Phi_0 = \frac{1}{\rho(\rho-1)} n_A \Phi_0 \quad (1.15)$$

In the following we show that, in the very strong *E*-field case, fractional filling factors may also occur, by variation of electron number *N* with *E* and *B* fixed. [In that case, the analogous width of the corresponding plateau is expected to be reduced, although a serious consideration of plateau-observability should include a study in the presence of disorder, a subject that is beyond the scope of the present article, or in the presence of edge states (see Section 2.5 for a qualitative discussion).]

## *2.1 High E-field strength*

When the electric field exceeds its first critical value, $E > \hbar \omega / e L_y$, inter L.L. overlap occurs. As *E* gets stronger, more and more L.Ls overlap and degenerate states that belong to different values of the quantum number *n* appear (in the previous case of a weak electric case, the standard Landau degeneracy had been completely lifted). Energy gap closes, and Fermi energy is always located on a single quantum state, with a significant number of available nearby states. Although Fermi energy is no longer in an energy gap, it will make jumps from one L.L to another by varying the magnetic field (or the particle number *N*). There are some critical values of *B* (the transition points) where a jump occurs, at which the Hall conductivity takes fractional values, as we shall see.

We now assume a general strength *E* field determined by $E = z \hbar \omega / e L_y$, (2.1)

with $z$ a continuous number that describes "overlap percentage": $j \leq z < j+1$ with $j = 0,1,2...$ . When $j = 0$, or $z \leq 1$, overlap vanishes and L.Ls can be filled independently, with the usual energy gap restored. This is just the case discussed in previous section. Eq. (2.1) can actually be derived for a certain electric field that obeys the following relation:

$$\varepsilon(n=j, Y_0 = -L_y/2) \leq \varepsilon(n=0, Y_0 = L_y/2) \leq \varepsilon(n=j+1, Y_0 = -L_y/2),$$

i.e. the electric field has a strength such that the single particle energy of the last electron $(Y_0 = L_y/2)$ in L.L. $n=0$ is greater than single particle energy of the first particle $(Y_0 = -L_y/2)$ in L.L. $n = j$ and lower than the single particle energy of the first particle in L.L. $n = j+1$. Then,

$$\hbar\omega(j+1/2) - eEL_y/2 \leq \hbar\omega/2 + eEL_y/2 \leq \hbar\omega(j+3/2) - eEL_y/2$$

that concludes in

$$j\frac{\hbar\omega}{eL_y} \leq E \leq (j+1)\frac{\hbar\omega}{eL_y}, \text{ or } E = z\frac{\hbar\omega}{eL_y}, \text{ with } j \leq z < j+1,$$

which is indeed (2.1).

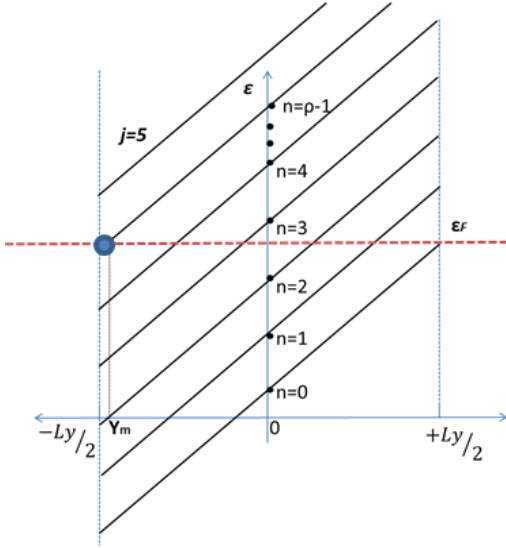

*Fig. 2.1: An example of different L.L mixture with **j=5** and **z=5.1** a possible position of Fermi Level (**ρ=6**), with only one full L.L. occupied and five L.Ls partially occupied. Ground state demands that the chemical potential of each distinct L.L (highest single-electron energy) must be equal, resulting in a purely horizontal Fermi energy.*

Now we procced to occupancies: For a constant $E$ and $B$ field (i.e. a constant energetic configuration) varying electron number $N$ results in variation of Fermi energy and a corresponding variation of occupied L.Ls. The goal here is to examine under what $N$-variations the Fermi level remains at a single (the topmost) L.L level. To achieve this, we fix Fermi level at the highest occupied L.L, $n=\rho-1$, $\rho \geq 1$, (see Fig. 2.1 above) with energy

$$\varepsilon_F = \hbar\omega(\rho - 1/2) + mV_D^2/2 + eEY_m, \quad (2.2)$$

with $Y_m$ the guiding center position of the last electron (with highest single particle energy) located at $n=\rho-1$ (see the isolated dot in Fig. 2.1). In this way, we ensure that variations of $N$ will result in moving $Y_m$ in the interval $Y_L \leq Y_m \leq Y_R$, with $Y_L = -L_y/2$, the left edge of $n=\rho-1$ L.L. and $Y_R$ is a critical guiding center above which L.L. $n=\rho$ is occupied. This can be determined by equating the Fermi energy to the single particle energy $\varepsilon(n=\rho, Y_0 = -L_y/2)$:

$$\hbar\omega(\rho - 1/2) + mV_D^2/2 + eEY_R = \hbar\omega(\rho + 1/2) + mV_D^2/2 - eEL_y/2,$$

resulting in: $Y_R = -L_y/2 + \hbar\omega/eE$, or, using eq. (2.1): $Y_R = -L_y/2 + L_y/z$. Any number $N$ that takes $Y_R$ above this critical value will result in a nonzero occupation of $n=\rho$ L.L. Therefore, to ensure that exactly $\rho$ L.Ls are occupied, $Y_m$ must vary in the following window:

$$-L_y/2 \leq Y_m \leq -L_y/2 + L_y/z, \quad (2.3)$$

guaranteeing that exactly $\rho$ L.Ls are occupied. Naturally this equation becomes pathological when $z$ is smaller than unity – the weak electric field case – $Y_R$ becomes infinite when $z \to 0$. This of course happens because the above equation is not valid in this case, where no inter-L.L overlaps occur. Limitations must therefore be imposed here. The smallest value that $z$ can take, must be above 1, with $z=1$ defining the non-overlap limit. If $z<1$ the above relation just becomes:

$$-L_y/2 \leq Y_m \leq L_y/2 \quad (2.4)$$

as earlier. Having now defined the Fermi level's exact location analytically, we proceed with our method by counting how many L.Ls are fully occupied and how many are partially occupied. An L.L. is completely filled with electrons only when the Fermi energy is greater than the single electron energy located at $Y_0 = L_y/2$ in each L.L. Using this information, we may write down the number of fully occupied L.Ls $(i_F)$ as a sum of theta functions over all L.Ls:

$$i_F(Y_m, \rho, z) = \sum_{n=0}^{+\infty} \theta\left[\varepsilon_F - \varepsilon\left(n, \frac{L_y}{2}\right)\right] = \sum_{n=0}^{+\infty} \theta\left[\rho - 1 - n + z\left(\frac{Y_m}{L_y} - \frac{1}{2}\right)\right] \text{ for } z > 1, \quad (2.5)$$

where $\theta[x]$ is a step function, $\theta[x]=1$ if $x \geq 0$ and $\theta[x]=0$ otherwise. Variations of $Y_m$ (i.e. for fixed $E$ and $B$) according to eq. (2.5) result in variations in $i_F$ as follows:

$$\underbrace{i_F^L(\rho, z)}_{Y_m = Y_L} \leq i_F(Y_m, \rho, z) \leq \underbrace{i_F^R(\rho, z)}_{Y_m = Y_R}, \quad (2.6)$$

with $i_F^L(\rho,z) = \sum_{n=0}^{+\infty} \theta[\rho-1-z-n]$ and $i_F^R(\rho,z) = \sum_{n=0}^{+\infty} \theta[\rho-z-n]$ are the numbers of completely filled L.Ls calculated at the edges of eq. (2.6). Let's see now some examples:

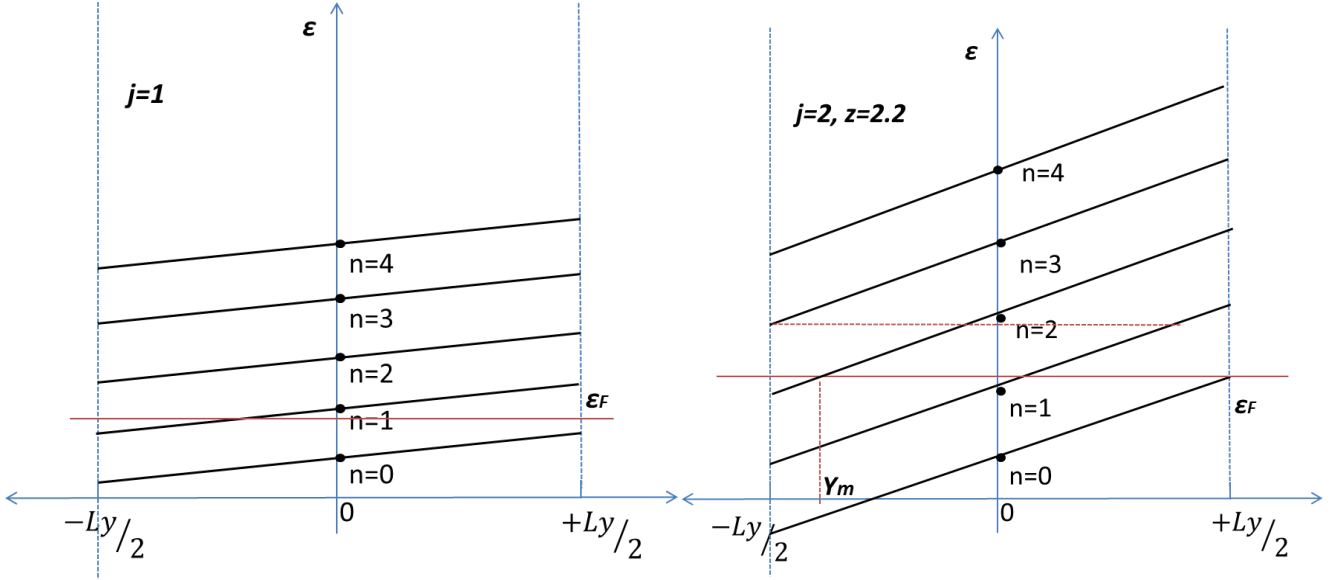

*Fig. 2.2*                      *Fig. 2.3*

In Fig. 2.2, $z$ is equal to unity, $z=1$ and $\rho=2$, then $i_F^L = \sum_{n=0}^{+\infty} \theta[-n] = \theta[0] = 1$, and $i_F^R = \sum_{n=0}^{+\infty} \theta[1-n] = \theta[1] + \theta[0] = 2$. In Fig. 2.3, $z=2.2$ and $\rho=3$, $i_F^L = 0$ and $i_F^R = 1$. It is interesting to note that for any value of the variable $z$, $i_F^L$ and $i_F^R$ always differ by one, which means that varying $N$ or equivalently, $Y_m$, in the interval defined by (2.4), one full L.L. is added to the system. This resembles the $E=0$ case, where variations of $Y_m$ in the same interval (but for $z=1$) the Fermi energy passes through all available states in the last L.L. until it reaches the next L.L., where an extra full L.L. increases the number $i_F$ by 1 (see Fig. 2.4 below). [This is actually consistent with the well-known corresponding result in the case we fold the system (in the $x$–direction) into a cylinder, that is a key result in the Laughlin argument [4] that gives the integral quantization of the Hall conductivity from general gauge arguments – or equivalently to the charge-pumping picture of Thouless [5] or a more general property of the so-called spectral flow [6] in topologically nontrivial systems, such as topological insulators [7].] It is interesting to see that when $z$ reaches its lowest possible value, $z=1$, $i_F^L = \rho-1$, meaning that the maximum number of partially occupied L.Ls is always 1. Needless to say that *z* remains a constant only when *E* and *B* fields are constant too, (or in a special case that both fields vary in the exactly same rate) and the only variable is the electron number. This happens because of eq. (2.1), $E = z\hbar\omega/eL_y$, which relates *B* to *E* through the variable *z*. For a constant electric field, varying *B* will result in a variation of *z* such that *E* remains a constant. The same principle holds for keeping *B* constant and varying *E*.

Generally, from definition of $i_F^R$ and $i_F^L$ we have that: $i_F^R = \rho - Int[z] + \delta$ and $i_F^L = \rho - Int[z] - 1 + \delta$, where $\delta = 1$ if *z* is an integer and $\delta = 0$ otherwise. These equalities hold for $\rho > Int[z] - \delta$. If $\rho = Int[z] - \delta$ or lower, then $i_F^R = 0$ and $i_F^L = 0$.

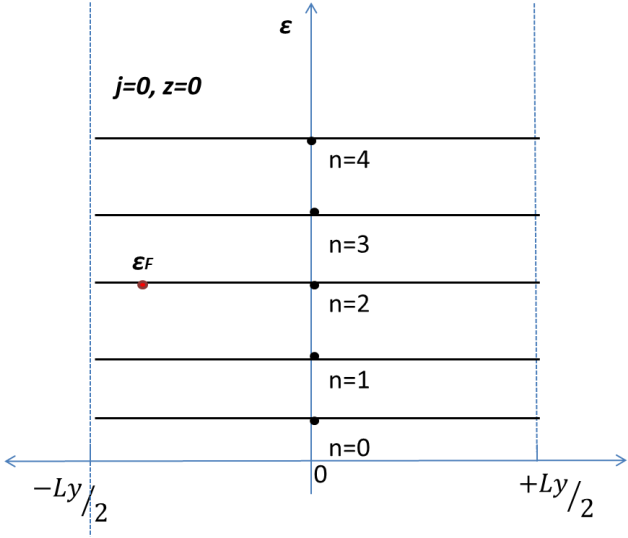

*Fig. 2.4:* For z=0 case (identical to $z \leq 1$ case), with no E-field, $i_F^L = \rho - 1$ and $i_F^R = \rho$. Left figure shows ρ=3 occupied L.Ls, with two of them fully occupied and the last one partially occupied. An addition of electrons will stimulate Fermi energy to pass through all available states in its right side until right end of L.L. n=2, where a full L.L. will be added to $i_F$.

Now, about partially filled L.Ls, these must intersect at a certain $Y_0$ Fermi energy. Optimal energy requires that states (per L.L.) starting with $Y_0 = -L_y/2$ until the intersection with the Fermi energy must be filled, while states with higher $Y_0$ s must be empty. These intersections can be easily found at the following points:

$$Y_0(l) = \frac{1}{eE}\left[\varepsilon_F - \hbar\omega\left(i_F + l + \frac{1}{2}\right) - \frac{1}{2}mV_D^2\right], \text{ with } l = 0,1,2...\rho - 1 - i_F, \quad (2.7)$$

where $\rho - i_F$ is the number of partially occupied L.Ls.

## 2.2 Number of states under Fermi energy

Each L.L contains $\Phi/\Phi_0$ available quantum states. Each state corresponds to a single spinless electron, in accordance to Pauli Exclusion Principle. The number of states under Fermi energy is a sum of $i_F$ full states and a sum term that counts all states in the partially occupied L.Ls from $-L_y/2$ until the intersection with Fermi energy:

$$\# states = i_F \frac{\Phi}{\Phi_0} + \sum_{l=0}^{\rho-1-i_F}\left[\frac{L_x B \varepsilon_F}{Ehc} - \frac{\omega B L_x}{2\pi cE}\left(i_F + l + \frac{1}{2}\right) - \frac{mV_D^2 L_x B}{2Ehc} + \frac{\Phi}{2\Phi_0}\right] \quad (2.8)$$

This number, in a canonical ensemble, is exactly equal to the constant particle number $N$:

$$N = i_F \frac{\Phi}{\Phi_0} + (\rho - i_F)\left[\frac{\varepsilon_F}{E}\frac{BL_x}{ch} - mc\frac{L_x}{h}\frac{E}{2B} + \frac{eBL_x}{ch}\frac{L_y}{2} - \frac{eB^2}{4\pi mE}\frac{L_x}{c^2}[\rho + i_F]\right] \quad (2.9)$$

$$N = i_F \frac{\Phi}{\Phi_0} + (\rho - i_F)\left[\frac{1}{4\pi E}\frac{eB^2 L_x}{mc^2}(\rho - 1 - i_F) + \frac{BL_x}{ch}eY_m + \frac{eBL_x}{ch}\frac{L_y}{2}\right]$$

with $i_F = i_F(\varepsilon_F, E, B)$ also a function of $\varepsilon_F$. This equation determines particle number $N$ when $\varepsilon_F$ is also known. Unfortunately, it is rather difficult to solve directly with respect to $\varepsilon_F$, but it can be easily determined numerically. Further simplification of (2.9) can be made using the window of values of $\varepsilon_F$, given by eq.(2.2) and (2.3):

$$\hbar\omega\left(\rho - \frac{1}{2}\right) + \frac{mV_D^2}{2} - eE\frac{L_y}{2} \leq \varepsilon_F \leq \hbar\omega\left(\rho - \frac{1}{2}\right) + \frac{mV_D^2}{2} + eE\left(-\frac{L_y}{2} + \frac{L_y}{z}\right), \quad (2.10)$$

which results in the following window of values for $N$:

$$\frac{B^2 L_x}{Emc^2 4\pi}e(\rho - i_F^L)[\rho - 1 - i_F^L] + \frac{\Phi}{\Phi_0}i_F^L \leq N \leq \frac{B^2 eL_x}{4\pi Emc^2}(\rho - i_F^R)[\rho - 1 - i_F^R] + \frac{\Phi}{\Phi_0}\left[i_F^R + \frac{(\rho - i_F^R)}{z}\right]$$

$$(2.11)$$

with $i_F^L = i_F^L(\rho, E, B)$ and $i_F^R = i_F^R(\rho, E, B)$. The above relation defines windows of values for $B$ (if it is considered as a variable) for constant $E$ field, and constant electron number $N$. Equivalently, one may prepare an experiment, where electron number is not conserved (i.e. an electric circuit) and keep $E$ and $B$ fixed. Note that the above relation shrinks to eq. (1.14) when $z = 1$:

$$\rho\frac{\Phi}{\Phi_0} \geq N \geq (\rho - 1)\frac{\Phi}{\Phi_0} \quad (2.12)$$

## 2.3 Thermodynamics

We now proceed to ground state energy calculation, which is a sum over all single electron states, until reaching the Fermi energy:

$$E = \frac{BL_x}{\Phi_0} \sum_{n=0}^{i_F-1} \int_{-L_y/2}^{L_y/2} dY_0 \left[ \hbar\omega\left(n+\tfrac{1}{2}\right) + \frac{1}{2}mV_D^2 + eEY_0 \right] + \frac{BL_x}{\Phi_0} \sum_{l=0}^{\rho-1-i_F} \int_{-L_y/2}^{Y_0(l)} dY_0 \left[ \hbar\omega\left(i_F+l+\tfrac{1}{2}\right) + \frac{1}{2}mV_D^2 + eEY_0 \right]$$

The left term describes the energy due to fully occupied L.Ls, while the right term describes the energy of partially occupied L.Ls. After a number of algebraic manipulations, and using eq. (2.7) we conclude to the following result for the total energy per electron:

$$\frac{E}{N} = -\frac{e^2 B^2}{4\pi mc^2 n_A} i_F \rho + \frac{e\hbar B}{2cm}(\rho + i_F) + \frac{L_y}{2(\rho - i_F)} \frac{En_A ch}{B} - eEL_y \left(\frac{\rho + i_F}{2(\rho - i_F)}\right) + \frac{mc^2 E^2}{2B^2}$$

$$+ \left[\frac{\rho i_F}{2(\rho - i_F)}\right] \frac{e^2}{hcn_A} L_y EB - \frac{e^2 \hbar B^3}{48\pi n_A m^2 c^3 EL_y}(\rho - i_F - 1)(\rho - i_F)(\rho - i_F + 1)$$

**(2.13)**

that consists of a sum of several terms which are not all symmetric (with respect to $E$ and $B$). The last term vanishes in weak $E$ field limit ($i_F = \rho - 1$). Note the interesting fact that the presence of the coupling term $EB$ requires fully occupied L.Ls, i.e. $i_F \neq 0$. This is a probably expected pattern resulting from axionic considerations in an electromagnetic field where the coefficient of coupling term involves fully occupied L.Ls (for the weak E field case, the corresponding coefficient is related to the Hall conductivity, and it was briefly mentioned in the last section, originating from the $EB$ – coupling term).

When $i_F = 0$, i.e. when all L.Ls are partially occupied ($\rho \leq Int[z] - \delta$), we have (from (2.13)):

$$\frac{E}{N} = \frac{e\hbar B}{2cm}\rho + \frac{L_y}{2\rho}\frac{En_A ch}{B} - \frac{eEL_y}{2} + \frac{mc^2 E^2}{2B^2} - \frac{e^2 \hbar B^3}{48\pi n_A m^2 c^3 EL_y}\rho(\rho^2 - 1)$$

and from (2.11):

$$\frac{B^2 eL_x}{Emc^2 4\pi}\rho(\rho - 1) \leq N \leq \frac{B^2 eL_x}{4\pi Emc^2}\rho(\rho - 1) + \frac{\Phi}{\Phi_0}\frac{\rho}{z}$$

Or, substituting $E = z\hbar\omega/eL_y$ we get

$$\frac{\rho}{2z}\frac{\Phi}{\Phi_0}(\rho - 1) \leq N \leq \frac{\rho}{2z}\frac{\Phi}{\Phi_0}(\rho + 1) \quad \textbf{(2.14)}$$

i.e. there are no quadratic terms with respect to $B$ appearing in the total energy, and there are no coupling terms either. Also, when the Fermi energy travels from one edge of n=ρ-1 to the other, it passes exactly through $\rho\Phi/z\Phi_0 < \Phi/\Phi_0$ states. On the other hand, when $\rho > Int[z] - \delta$, i.e. $i_F \neq 0$, the Fermi energy passes through exactly $\Phi/\Phi_0$ states.

Now, we want to plot eq. (2.13) by keeping $E$ and $B$ fixed and vary $N$ in the interval given by (2.11). Using $E$ as in (2.1) we rewrite (2.13) and (2.11) as

$$\frac{E}{\frac{e^2 B^2 L_x L_y}{4\pi mc^2}} = -i_F \rho + \frac{N\Phi_0}{\Phi}(\rho + i_F) + \frac{zN^2}{(\rho - i_F)}\frac{\Phi_0^2}{\Phi^2} - z\frac{N\Phi_0}{\Phi}\left(\frac{\rho + i_F}{\rho - i_F}\right) + Nz^2 \frac{\Phi_0^2}{\Phi^2}$$

$$+ \frac{z\rho i_F}{\rho - i_F} - \frac{1}{12z}(\rho - i_F - 1)(\rho - i_F)(\rho - i_F + 1)$$

and $\quad \frac{1}{2z}(\rho - i_F^L)\left[\rho - 1 - i_F^L\right] + i_F^L \leq N\frac{\Phi_0}{\Phi} \leq \frac{1}{2z}(\rho - i_F^R)\left[\rho + 1 - i_F^R\right] + i_F^R$

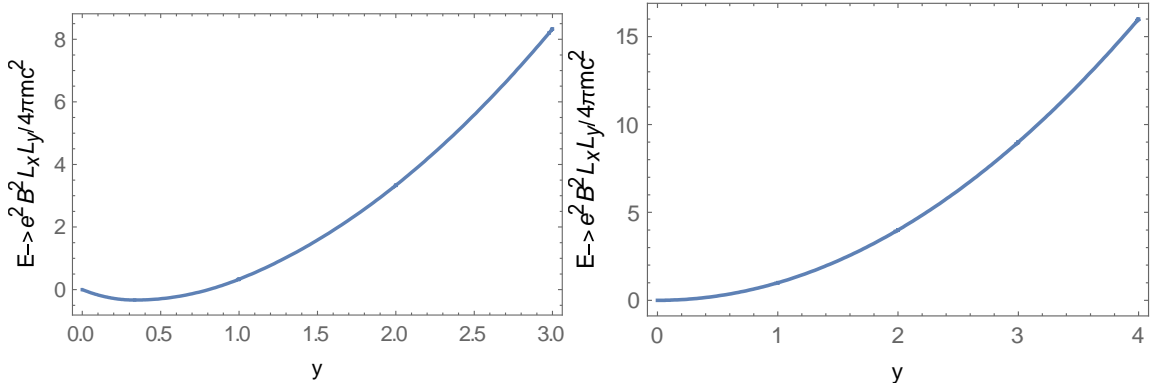

***Fig. 2.5.*** *Total energy (in units of $e^2 B^2 L_x L_y / 4\pi mc^2$) as function of particle number (in units of $\Phi/\Phi_0$, $y = N\Phi_0/\Phi$) for z=3 (Left), z=1 (Right, low E field limit).*

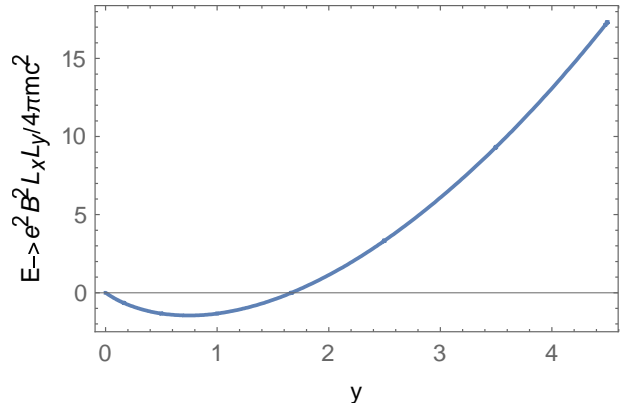

***Fig. 2.6.*** *Total energy (in units of $e^2 B^2 L_x L_y / 4\pi mc^2$) as function of particle number (in units of $\Phi/\Phi_0$, $y = N\Phi_0/\Phi$) for z=6.1.*

A few remarks are in order about these figures: It is clear that when the E field is exceedingly strong, in contrast to the case of a weak E field, there appears a global minimum with respect to particle number variations. This minimum is a consequence of the E field effect on the thermodynamic properties; as E gets stronger, more and more states will gain negative energy, (see for example Fig. 2.3). As these states get occupied by electrons, the total energy will become negative at first, then will rise up to positive values, because states with positive energy will begin to fill up. The result of this competition is this minimum, which can be fully controlled by examining the corresponding z-value. The greater z gets, the more negative-energy states will be occupied, and the minimum will move to greater particle numbers (or larger particle densities). In comparison with the low E field case, where the positive states are more, in this case the total energy may be even lower.

We now plot the total energy as a function of variable z, for a constant E field with $y = eEL_y / \varepsilon_F$ (a suitable measure for the E field in units of 2D Fermi energy in the E=0, B=0 case) and $B = mcL_y E / \hbar z$ from eq. (2.1):

$$\frac{E}{N\varepsilon_F} = -\frac{y^2}{2z^2}i_F \rho + \frac{y}{2z}(\rho + i_F) + \frac{1}{(\rho - i_F)}\frac{z}{2} - \frac{y}{2}\left(\frac{\rho + i_F}{\rho - i_F}\right) + \frac{z^2}{L_y^2 4\pi n_A} + \left[\frac{\rho i_F}{2(\rho - i_F)}\right]\frac{y^2}{z}$$
$$- \frac{1}{24}\frac{y^2}{z^3}(\rho - i_F - 1)(\rho - i_F)(\rho - i_F + 1)$$

where z lies in the following window:

$$\frac{1}{2z^2}(\rho - i_F^L)[\rho - 1 - i_F^L] + \frac{1}{z}i_F^L \leq \frac{1}{y} \leq \frac{1}{2z^2}(\rho - i_F^R)[\rho + 1 - i_F^R] + \frac{1}{z}i_F^R$$

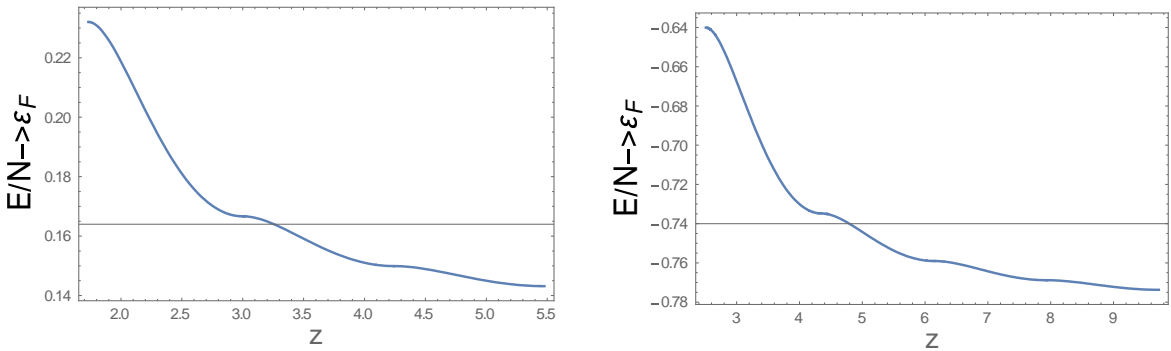

*Fig. 2.7. Total energy per particle (in units of Fermi energy in the absence of E and B) as function of z (analogous to inverse B) for y=3 (Left), y=6.3 (Right, low E field limit).*

We should point out here that although the energy is periodic function with respect to inverse $B$ (this is the de Haas-van Alphen effect [8] – see how this can be derived at $T=0$ in various cases from rather elementary considerations in ref. [9]), the windows of values of $z$ (or $1/B$) seem to be influenced by the presence of the electric field, namely $\delta(1/B) = \hbar/mcL_y E$, or, by using $y = eEL_y/\varepsilon_F$ we find that $\delta(1/B) = \hbar e/ymc\varepsilon_F = 1/yn_A\Phi_0$ **(2.15)** which deviates from the standard semiclassical periodicities $\delta(1/B) = 1/n_A\Phi_0$ for a 2D system [9].

In addition, it should be pointed out that the correctness of our results is witnessed by the fact that the total energy turns out to be a smooth (continuous and differentiable) function of $z$ for every $z$ (i.e. the positions of the windows match with the corresponding expressions) – this smoothness of the figures (in their joining through different window-values) strictly testifying for the *analytical correctness of the overall expressions* that we derived for the total energy.

## *2.4 Hall conductivity*

The Hall conductivity is defined as: $\sigma_H = en_A c/B$ **(2.16)** and is usually plotted as a function of magnetic field $B$. For convenience we will make here use of eq. (2.11) and examine $\sigma_H$ as a function of electron number $N$ instead of $B$. Starting therefore from

$$\frac{B^2 L_x}{Emc^2 4\pi} e(\rho - i_F^L)[\rho - 1 - i_F^L] + \frac{\Phi}{\Phi_0} i_F^L \leq N \leq \frac{B^2 eL_x}{4\pi Emc^2}(\rho - i_F^R)[\rho - 1 - i_F^R] + \frac{\Phi}{\Phi_0}\left[i_F^R + \frac{(\rho - i_F^R)}{z}\right]$$

**(2.17)**

we will show that by varying $N$, fractional filling factors appear, with no interactions and no impurities taken into account. $E$ and $B$ fields are considered as constants (and therefore so is $z$) throughout all $N$ (and $\rho$) variations. Substituting eq. (2.1) into (2.17) we may eliminate $E$:

$$\frac{\Phi}{\Phi_0}\left[\frac{1}{2z}(\rho - i_F^L)(\rho - i_F^L - 1) + i_F^L\right] \leq N \leq \frac{\Phi}{\Phi_0}\left[\frac{1}{2z}(\rho - i_F^R)(\rho - i_F^R + 1) + i_F^R\right]$$

Because for all numbers $\rho$ and $i_F^R$ the two edges of (2.17) differ by $\dfrac{\Phi}{\Phi_0}$, we choose to calculate $\sigma_H$ only at the special values of electron number $N$ given by the maximum edge of (2.17):

$$N = \frac{\Phi}{\Phi_0}\left[\frac{1}{2z}(\rho - i_F^R)(\rho - i_F^R + 1) + i_F^R\right],$$

and $\sigma_H$ is then:

$$\sigma_H = \frac{en_A c}{B} = \frac{e^2}{h}\left[\frac{1}{2z}(\rho - i_F^R)(\rho - i_F^R + 1) + i_F^R\right] \text{ with } i_F^R(\rho,z) = \sum_{n=0}^{+\infty}\theta[\rho - z - n]$$

Let's examine now some cases regarding *z*-values:

### *z is an integer:*

1) $z=1 \Rightarrow i_F^R = \rho : \sigma_H = \frac{\rho e^2}{h}$

2) $z>\rho: \Rightarrow i_F^R = 0: \sigma_H = \frac{e^2}{h}\left[\frac{1}{2z}\rho(\rho+1)\right] = \frac{e^2}{h}\frac{1}{z}, \frac{e^2}{h}\frac{3}{z}, \frac{e^2}{h}\frac{6}{z}, \frac{e^2}{h}\frac{10}{z}.....$

3) $z=\rho: \Rightarrow i_F^R = 1: \sigma_H = \frac{e^2}{h}\left[\frac{1}{2}(z-1)+1\right]$.

4) $z<\rho: \Rightarrow i_F^R = \rho - z + 1, \sigma_H = \frac{e^2}{h}\left[\frac{1}{2}(z-1)+1\right]$.

$$\sigma_H = \frac{e^2}{h}\left[\rho - \frac{1}{2}(z-1)\right] = \frac{e^2}{2h}[z+3], \frac{e^2}{2h}[z+5], \frac{e^2}{2h}[z+7]...$$

More compactly, for a fixed integer *z*, varying *ρ* will result in the following values for $\sigma_H$:

$$\sigma_H = \underbrace{\frac{e^2}{h}\frac{1}{z}, \frac{e^2}{h}\frac{3}{z}, \frac{e^2}{h}\frac{6}{z}, \frac{e^2}{h}\frac{10}{z}...}_{z-1 Terms} \frac{e^2}{h}\left[\frac{1}{2}(z-1)+1\right], \frac{e^2}{2h}[z+3], \frac{e^2}{2h}[z+5], \frac{e^2}{2h}[z+7]...$$

**(2.18)**

where, for *z*=1 we get the known results: $\sigma_H = \frac{e^2}{h}, \frac{2e^2}{h}, \frac{3e^2}{h}, \frac{4e^2}{h}...$

### *z is not an integer*

For a non-integer z we have:

$$\sigma_H = \underbrace{\frac{e^2}{zh}, \frac{3e^2}{zh}, \frac{6e^2}{zh}, \frac{10e^2}{zh},...}_{jTerms} \frac{e^2}{h}\left[1+\frac{j}{2z}(j+1)\right], \frac{e^2}{h}\left[2+\frac{j}{2z}(j+1)\right].. , \textbf{(2.19)}$$

where $j = Int(z)$.

We re-emphasize here that the above fractional values in Hall conductivity do not result from any interactions (i.e. they are not related to the Physics of the Fractional Quantum Hall Effect) but are a genuine consequence of a high *E*-field strength that causes the L.L overlaps.

## 2.5 Plateaux formation

When the external *E* field is relatively small, i.e. given by eq. (1.3), the Hall conductance has a plateau–like structure. To see this, imagine that we vary in a continuous matter the particle number, keeping *B* fixed. As a result, the Fermi energy is continuously moved from the beginning of a Landau Level to the right end, after passing through all states in that L.L. And when Fermi energy is in the interior of a L.L., Hall conductivity remains a constant *and it only changes at the critical N values* ($N = \rho\Phi/\Phi_0$), namely when Fermi energy has a transition between adjacent L.Ls.

In the strong *E* field case however, this is not the case. When Fermi energy lies in the interior of the $\rho^{th}$ L.L, plateaux are destroyed, because the energy spectrum is no longer discretized. It is as if one has a single L.L. band (all L.Ls form a single band), with an infinite number of neighboring states to accommodate all electrons (a metallic character) and Hall conductivity is more likely to behave as in the classical case, with an eternal increasing (or decreasing) character by varying either *N* or *B*.

Inclusion of impurity potentials [10] however may change this expectation in the sense that, there might be a case where Fermi energy will lie in a mobility gap (which is the true criterion for the existence of QHE), and some of the extended states per L.L might be occupied, resulting in a blurred quantization of Hall conductance. This is a matter that needs to be investigated more carefully. In this case, the previous $\sigma_H$ values will correspond to certain plateaux by varying *B* or *N*.

## 2.6 Inclusion of edge states

In real-life samples, however, electronic systems are confined, and the confinement and impurity potential must be seriously taken into account. QHE manifestation is accomplished by considering both effects on transport properties (the confinement and disorder). In this case, a thermodynamic approach may be of low importance and insufficient to explain the robust quantization of Hall conductivity. Here, we will try to visualize a special case, where the electric field is strong enough, so that it causes an inter-L.L overlap and the system is at the same time confined in *y*-direction. The key to solve this problem lies in the range of influence of the confinement potential. As one starts from *y*=0, and moving towards lets say, the right end ($L_y/2$) he will experience the confinement potential influence only on a certain distance from the end (a few multiples of magnetic length $l_B$). From that point on, L.Ls will start to rise abruptly.

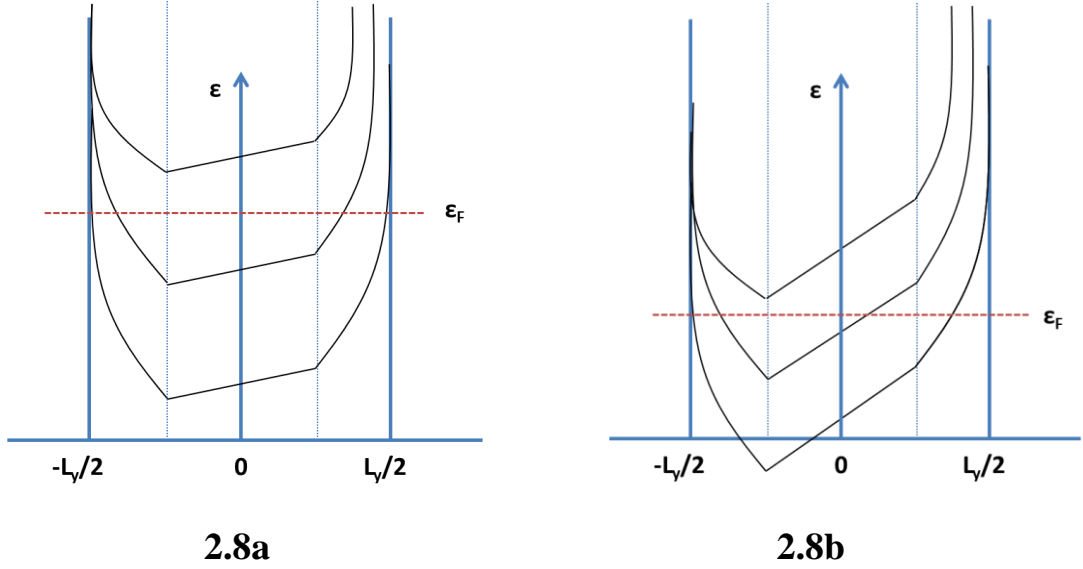

**2.8a**  **2.8b**

*Fig. 2.8. Impact of the edges on Landau Levels. (*a*) low-E field, (*b*) strong-E field case*

Now, if the electric field is low enough (Fig. 2.8a) there is no inter-L.L overlap and the Fermi energy will always intersect the same number of edge states at the left end and right end respectively. But in the case of a strong *E* field (Fig. 2.8b), things may change dramatically; if (for a certain *E* field) the overlap starts at a point before the confinement potential influence, there might be a case where the Fermi energy intersects two points (two L.Ls) at the left end and only one point (L.L) at the right end (where the edge states are located). This is possible only if there is an additional intersection in the bulk of a L.L. (see the middle L.L at Fig. 2.8b). Varying then the Fermi energy will not change the number of intersections, until the Fermi energy reaches the bottom of the top L.L. and so on. In other words, the occupations of edge states will remain the same, until another L.L will be occupied. Therefore, *there might be some robustness of the Hall values (found above)* before this crossover occurs.

## 3.  Generalization to Relativistic systems

### 3.1 Graphene – Low E-field strength

Our results can also be applied to Relativistic solid state systems that obey a Dirac-type of equation, such as Graphene [11] and topological insulators [7]. In this Section we give a first analytical study of strong electric field effects in the ground state energetics and also in transport properties of Graphene, i.e. Hall conductivity. We find again that, when the *E*-field is strong enough, L.Ls with different quantum numbers overlap, and the Hall conductivity becomes electric and magnetic field–dependent, indicating the non-linearity of Hall conductivity in relativistic systems, this having more refined consequences compared to the previous nonrelativistic case**;** we obtain, for example, the possibility of irrational values (the observability of which is of course up to the uncertainties related to the lack of disorder, already mentioned in the last section).

We start with Graphene's energy spectrum when placed in homogeneous crossed electric $\vec{E} = E\hat{x}$ and magnetic $\vec{B} = B\hat{z}$ fields, considering only the positive branch [12] (we continue in this section with SI units):

$$\varepsilon_{n,k_y} = \sqrt{2n\hbar eB}u_F\left(1-\beta^2\right)^{3/4} + \hbar u_F \beta k_y, \quad (3.1)$$

where $\beta = E/u_F B$ is a dimensionless parameter smaller than unity and

$$X_0 = l_B^2 k_y - \text{sgn}(n)\frac{\sqrt{2|n|}l_B \beta}{\left(1-\beta^2\right)^{1/4}} \quad (3.2)$$

is the guiding center eigenvalue, and $l_B = \sqrt{\hbar/eB}$ is the magnetic length. Each L.L (indexed by $n$) has $\Phi/\Phi_0$ available states (independent of $E$ strength) and may host up to $4\Phi/\Phi_0$ electrons, due to spin (a factor of two) and valley degeneracy (another factor of two) except for $n=0$ L.L., which has capacity $2\Phi/\Phi_0$, because it is equally shared with holes.

Using (3.2) we may rewrite (3.1) as:

$$\varepsilon_{n,k_y} = \frac{\sqrt{2n\hbar eB}u_F}{\left(1-\beta^2\right)^{1/4}} + eEX_0 \quad, n=0,1,2.. \quad (3.3)$$

with the restriction $-L_x/2 \leq X_0 \leq L_x/2$. Using (3.3), we determine the inter-L.L. energy gap as:

$$\delta\varepsilon_{n,k_y} = \frac{\sqrt{2\hbar eB}u_F}{\left(1-\beta^2\right)^{1/4}}\left[\sqrt{n+1}-\sqrt{n}\right] \quad (3.4)$$

This energy gap depends on *E*-field, in contrast to conventional semiconductor systems (of last section); in addition, the greater L.L. index, the smaller the gap. This means that in some L.L. range there will be unavoidable overlap accompanied with a zero energy gap (and the system is then metallic). This is exactly the case we will examine later. Now, for adjacent L.Ls, no overlap is observed only when the following condition holds:

$$\varepsilon(n,L_x/2) \leq \varepsilon(n+1,-L_x/2) \quad (3.5)$$

which leads to: $\dfrac{\sqrt{2e\hbar B}u_F}{\left(1-\beta^2\right)^{1/4}}\left[\sqrt{n+1}-\sqrt{n}\right]\geq eEL_x$ **(3.6).**

For small enough indexes *n*, and small enough *E*-fields, inequality (3.6) will be satisfied. But, for a given *E*-field, there will be a critical L.L. index $i_F$ above which (i.e. $n=i_F+1$) the inequality (3.6) changes sign:

$$\dfrac{\sqrt{2e\hbar B}u_F}{\left(1-\beta^2\right)^{1/4}}\left[\sqrt{i_F+2}-\sqrt{i_F+1}\right]\leq eEL_x \text{ (3.7)}$$

This $i_F$ describes the topmost L.L. that does not overlap with other L.Ls, while all remaining L.Ls indexed by $n=i_F+1, i_F+2...$ become continuous, without any energy gap and with always available neighboring states for an electron to be scattered in. It is useful to define a new continuous number *z*, which obeys:

$$\sqrt{i_F+1}-\sqrt{i_F}\geq z \geq \sqrt{i_F+2}-\sqrt{i_F+1} \text{ (3.8)}$$

This number, when put in (3.7), results in the following equality:

$$\dfrac{\sqrt{2e\hbar B}}{\left(1-\beta^2\right)^{1/4}}u_F z = eEL_x \text{ (3.9)}$$

which will be used later below. We now proceed to our calculation of thermodynamic properties by considering that Fermi energy is always located at an L.L. whose index ($n=\rho-1$, with $\rho=1,2,3..$) always obeys eq. (3.6), i.e. $\rho-1\leq i_F$. In this manner, all L.Ls are occupied independently, and when we have full occupations, Fermi energy will surely lie in an energy gap.

We make the assumption that electron number is such that $\rho$ L.Ls are occupied, with the last L.L. ($n=\rho-1$) partially occupied. Extra care must be taken in the counting of states, recalling the special case of the lowest L.L. $n=0$, which has capacity $2\Phi/\Phi_0$ electrons, while all other L.Ls may accommodate up to $4\Phi/\Phi_0$ electrons. Considering this, we find that the electron number lies in the following window of values (for a constant electromagnetic field):

$$(2\rho-3)\dfrac{2\Phi}{\Phi_0}\leq N \leq (2\rho-1)2\dfrac{\Phi}{\Phi_0} \text{ (3.10)}$$

(Note that when $\rho=1$, eq. (3.10) becomes: $0\leq N\leq 2\Phi/\Phi_0$). For a constant electron number *N*, we can solve (3.10) with respect to *B*:

$$\frac{1}{2(2\rho-3)}n_A\Phi_0 \geq B \geq \frac{1}{2(2\rho-1)}n_A\Phi_0 \quad \textbf{(3.11)}$$

What this inequality means is that when $B$ lies in the above range, the first $\rho-1$ ($n=0,1,2\ldots\rho-2$) L.Ls are fully occupied and the final L.L. ($n=\rho-1$) is partially occupied (it is filled with $N-(2\rho-3)2\Phi/\Phi_0$ electrons). Eq. (3.11) also describes the well-known unconventional Hall effect in graphene:

$$\sigma_H = \frac{n_A e}{B} = (2\rho-1)2\frac{\Phi}{L_x L_y \Phi_0 B}e = (2\rho-1)2\frac{e^2}{h} = \left(\rho-\frac{1}{2}\right)\frac{4e^2}{h} \quad \textbf{(3.12)}$$

where the Hall conductivity is quantized in half integer multiples of the quantity $4e^2/h$.

We now calculate the total internal energy of the system, which has to be minimalized when the temperature is zero ($T=0$):

$$E = \sum_{n,X_0} \varepsilon_{n,k_y} = \sum_{n,X_0} \frac{\sqrt{2n\hbar eBu_F}}{(1-\beta^2)^{1/4}} + eEX_0 \quad \textbf{(3.13)}$$

Considering the following transformations:

$$\sum_{X_0} \to \frac{4BL_y}{\Phi_0} \int_{-L_x/2}^{L_x/2} dX_0 \text{ for } n>0 \text{ and } \sum_{X_0} \to \frac{2BL_y}{\Phi_0} \int_{-L_x/2}^{L_x/2} dX_0 \text{ for } n=0 \text{ L.L.} \quad \textbf{(3.14)}$$

we can write:

$$E = E_{FULL} + E_{PART}, \quad \textbf{(3.15)}$$

where $E_{FULL}$ is the energy of the fully occupied L.Ls and $E_{PART}$ is the energy of the last partially occupied L.L.

$$E_{FULL} = \frac{4BL_y}{\Phi_0}\sum_{n=1}^{\rho-2}\int_{-L_x/2}^{L_x/2} dX_0 \frac{\sqrt{2n\hbar eBu_F}}{(1-\beta^2)^{1/4}} + \frac{4BL_y eE}{\Phi_0}\sum_{n=1}^{\rho-2}\int_{-L_x/2}^{L_x/2} dX_0 X_0$$
$$= \frac{4\Phi}{\Phi_0}\frac{\sqrt{2\hbar eBu_F}}{(1-\beta^2)^{1/4}}\sum_{n=0}^{\rho-2}\sqrt{n} \quad \textbf{(3.16)}$$

To calculate $E_{PART}$, it is necessary to determine the last electron's guiding center position in L.L $n=\rho-1$.

$$X_0 = l_B^2 \frac{2\pi l}{L_y} - \frac{\sqrt{2(\rho-1)}l_B\beta}{(1-\beta^2)^{1/4}} \quad \textbf{(3.17)}$$

The index $l$, appearing in $k_y$, has a starting value $l_0$ that needs to be determined by the condition $X_0 = -L_x/2$, (left edge):

$$-\frac{\Phi}{2\Phi_0} + \frac{L_y}{2\pi l_B}\frac{\sqrt{2(\rho-1)}\beta}{(1-\beta^2)^{1/4}} = l_0 \quad (3.18)$$

But the last L.L hosts $\left[N-(2\rho-3)\frac{2\Phi}{\Phi_0}\right]/4$ states for the remaining electrons to be placed inside, so we conclude that the last electron in $n=\rho-1$ L.L. has a guiding center position:

$$X_{0MAX} = l_B^2 \frac{2\pi}{L_y}\left[\frac{N}{4} - \rho\frac{\Phi}{\Phi_0} + \frac{\Phi}{\Phi_0}\right] \quad (3.19)$$

For consistency reasons we check eq. (3.19) for certain values of particle number, i.e. when $N=(2\rho-3)2\Phi/\Phi_0$ then $X_0 = -L_x/2$ and when $N=(2\rho-1)2\Phi/\Phi_0$ we have $X_0 = L_x/2$. So far we are correct. The energy of the electrons located at the Fermi L.L. $E_{PART}$ is:

$$E_{PART} = \frac{4BL_y}{\Phi_0}\int_{-L_x/2}^{X_{0MAX}} dX_0 \frac{\sqrt{2(\rho-1)\hbar eB}u_F}{(1-\beta^2)^{1/4}} + eE\frac{4BL_y}{\Phi_0}\int_{-L_x/2}^{X_{0MAX}} dX_0 X_0$$

$$= \frac{4BL_y}{\Phi_0}\frac{\sqrt{2(\rho-1)\hbar eB}u_F}{(1-\beta^2)^{1/4}}[X_{0MAX} + L_x/2] + eE\frac{4BL_y}{\Phi_0}\left[\frac{X_{0MAX}^2}{2} - \frac{L_x^2}{4}\right] \quad (3.20)$$

Substituting eq. (3.19) in eq. (3.20) we get:

$$E_{PART} = \frac{\sqrt{2(\rho-1)\hbar eB}u_F}{(1-\beta^2)^{1/4}}\left[N - 2(2\rho-3)\frac{\Phi}{\Phi_0}\right]$$

$$+ \frac{hN^2E}{8BL_y} - eEN(\rho-1)L_x + e^2E\frac{2BL_y}{h}L_x^2\left[\rho^2 - 2\rho + \frac{1}{2}\right] \quad (3.21)$$

From eqs (3.16) and (3.21) we obtain the final result for the total energy per electron of the system:

$$\frac{\mathrm{E}_{TOT}}{N} = \frac{\varepsilon_F}{(1-\beta^2)^{1/4}} \left[ 4\left(\frac{B}{n_A \Phi_0}\right)^{3/2} \left( 2\sum_{n=1}^{\rho-2} \sqrt{n} - (2\rho-3)\sqrt{(\rho-1)} \right) + 2\left(\frac{B}{n_A \Phi_0}\right)^{1/2} \sqrt{(\rho-1)} \right]$$
$$+ \frac{n_A \Phi_0 eEL_x}{8B} - eE(\rho-1)L_x + eE\frac{2BL_x}{n_A \Phi_0}\left[\rho^2 - 2\rho + \frac{1}{2}\right]$$

(3.22),

with $\sum_{n=1}^{\rho-2} \sqrt{n} = -\zeta\left(-\frac{1}{2}, \rho-1\right) - \frac{\zeta\left(\frac{3}{2}\right)}{4\pi}$, where $\zeta(-1/2, \rho-1)$ is the Hurwitz zeta function and $\zeta(3/2)$ is the Riemann function. Note that the term proportional to the *E* field appearing in (3.22) when expressed per electron,

$$e^2 E \frac{2B}{n_A h} L_x \left[\rho^2 - 2\rho + \frac{1}{2}\right] + \frac{hNE}{8BL_y}, \quad (3.23)$$

gives the following magnetization (which is also proportional to *E*):

$$-e^2 E \frac{2}{n_A h} L_x \left[\rho^2 - 2\rho + \frac{1}{2}\right] + \frac{hNE}{8B^2 L_y} \quad (3.24)$$

When all $\rho$ L.Ls are fully occupied, then $B = n_A \Phi_0 / 2(2\rho-1)$, and this term becomes:

$$\frac{L_x}{2n_A}\left([4\rho-1]\frac{e^2}{h}\right)E \sim \frac{L_x}{2n_A}\frac{\sigma_H}{2}E \quad (3.25)$$

which is in accordance to the non - relativistic case (see corresponding term of (1.11)). Although this $\sigma_H = [4\rho-1]e^2/h$ differs from eq. (3.12) where $\sigma_H = (2\rho-1)2e^2/h$ this point needing further clarification. Let us then check some limits: When the electric field is absent, *E*=0, the total energy per electron becomes (in units of Fermi energy in the absence of *B*, $\varepsilon_F = \hbar u_F \sqrt{\pi n_A}$):

$$\frac{\mathrm{E}_{TOT}}{N} = \varepsilon_F \left[ 4\left(\frac{B}{n_A \Phi_0}\right)^{3/2} \left[ 2\sum_{n=1}^{\rho-2} \sqrt{n} - (2\rho-3)\sqrt{(\rho-1)} \right] + 2\left(\frac{B}{n_A \Phi_0}\right)^{1/2} \sqrt{(\rho-1)} \right] \quad (3.26)$$

This is the result we would have gotten had we solved the problem from the beginning without the *E*-field.

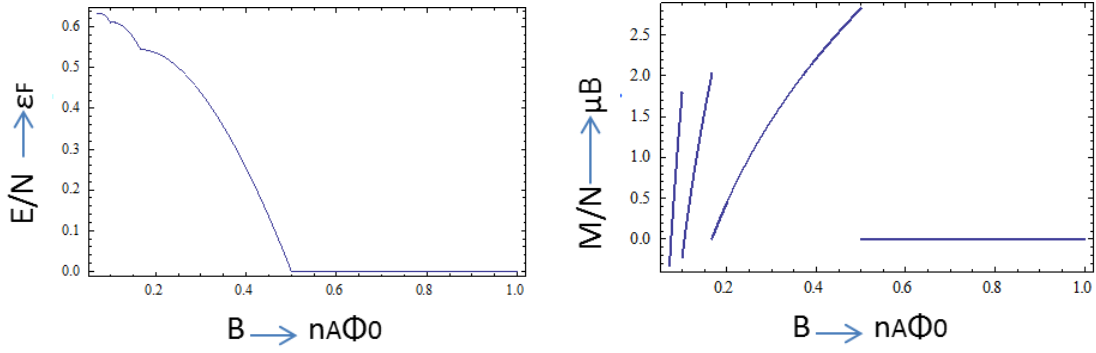

**Fig. 3.1:** *Energy in units of $\varepsilon_F$ and magnetization in units of Bohr magneton $\mu_B$ as functions of magnetic field in the absence of an electric field.*

When $B$ is at a critical value, $B = n_A \Phi_0 / 2(2\rho-1)$, (all $\rho$ L.Ls fully occupied) then:

$$E_{TOT} = 2N\hbar u_F \frac{1}{(2\rho-1)} \sqrt{\frac{n_A 2\pi}{(2\rho-1)}} \left[ \sum_{n=1}^{\rho-2} \sqrt{n} + \sqrt{(\rho-1)} \right]$$
$$= 2N\hbar u_F \sqrt{\frac{n_A 2\pi}{(2\rho-1)^3}} \sum_{n=1}^{\rho-2} \sqrt{n} + 2N\hbar u_F \sqrt{\frac{n_A 2\pi(\rho-1)}{(2\rho-1)^3}} \quad (3.27)$$

If $\rho=1$ then $E_{TOT} = 0$, as one would expected, because all electrons are placed at the lowest L.L. having zero energy.

If $\rho=2$ $E_{TOT} = \sqrt{2\hbar eB} u_F \left[ 4\frac{\Phi}{\Phi_0} \right]$ and so on. In all cases, we can clearly see the difference between the case of 2DEG in conventional semiconductors (note, for example, that at the critical values of $B$ the total energy is no longer equal to the Fermi energy but keeps rising up – see Fig. 3.1 and its differences with Fig. 1.2).

## 3.2 Stronger E-field

The inequality

$$\frac{\sqrt{2e\hbar B} u_F}{(1-\beta^2)^{1/4}} \left[ \sqrt{i_F+1} - \sqrt{i_F} \right] \geq eEL_x,$$

is the criterion where no overlap occurs, as long as $\rho-1 \leq i_F$. Now, if we have 'more' electrons and have to place them in available states, we will finally reach the condition $\rho > i_F + 1$. In this case, states indexed by $n = i_F + 2 ..... \rho-1$, overlap, and the energy gap closes at Fermi energy. This means that graphene gains a metallic *E*- field induced character and Hall conductivity is modified (analogous to the non-relativistic case) by a term containing both electric and magnetic field.

Recall from previous work [13] that the following relation also holds:

$$\beta < 1 \Rightarrow \frac{E}{u_F B} < 1 \Rightarrow E < u_F B,$$

(so that no collapse of L.Ls occurs), which can always be made true as long as we treat particle number as our variable, and keep $E$ and $B$-fields constant. Or, equivalently, we may keep $N$ and $B$ fixed and treat $E$ as our main variable, starting from zero until reaching maximum value $E_{MAX} = u_F B$. We remind here the reader of the limitations of this problem regarding also the magnetic field, which has to be such that the magnetic length is much greater than lattice constant (i.e. we will not study any Hofstadter effects in this work). Either way, our results and predictions can be obtained by satisfying all possible limitations. For example, we may prepare an experiment where $E$ is strong enough for inter L.L. overlap to occur, and at the same time $B$ is also strong enough to satisfy the above inequality.

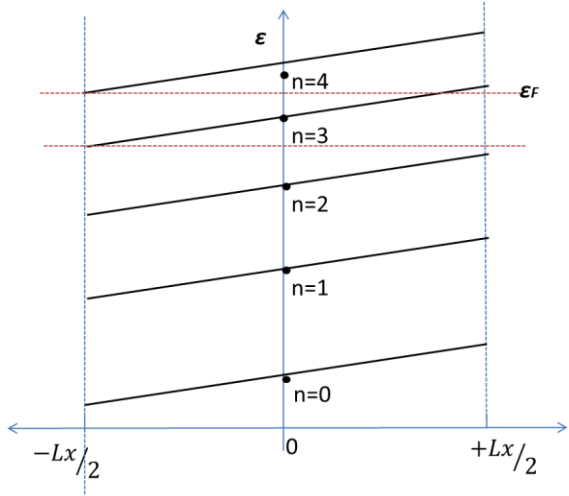

*Fig. 3.2: Schematic Representation of energy levels in graphene. Figure shows $\varepsilon_{n,k_y}$ vs $X_0$. The inter-L.L. energy gap gets smaller by increasing Landau Level index n. States n=0,1 and 2 do not overlap with each other, but states n=3,4,5… and so on, do. If the Fermi energy is located at n=0,1 or 2, we expect that Hall conductivity will still be quantized in the form $\sigma_H = (2\rho-1)2e^2/h$, with $\rho=1,2,$ and 3.*

We examine now the case appearing in Fig. 3.2. Landau levels $n=0,1$ and $2$ can be occupied independently, while further occupations above $n=2$ L.L. leads to the unavoidable mixing of states; this makes the problem interesting, in the sense that peculiarities may arise both in thermodynamic and transport properties. We fix Fermi energy $\varepsilon_F$ at L.L $n=\rho-1$:

$$\varepsilon_F = \frac{\sqrt{2(\rho-1)\hbar eB u_F}}{(1-\beta^2)^{1/4}} + eEX_{0M} \quad (3.28)$$

with $X_{0M}$ the guiding center position of the last electron in the Fermi level $n=\rho-1$. We consider now the simplest case: From Fig. (3.2), we will examine the case where all states are occupied up to the Fermi energy (eq. 3.28). The Fermi energy is located at $n=4$ with $X_{0M} = -L_x/2$ :

$$\varepsilon_F = \frac{\sqrt{8\hbar eB}u_F}{(1-\beta^2)^{1/4}} - \frac{eEL_x}{2} \quad (3.29)$$

To determine the number of states in L.L. $n=3$, *we examine its intersection with Fermi energy* (see Fig. (3.2)):

$$\frac{\sqrt{6\hbar eB}u_F}{(1-\beta^2)^{1/4}} + eEX_{0F} = \frac{\sqrt{8\hbar eB}u_F}{(1-\beta^2)^{1/4}} - \frac{eEL_x}{2} \Rightarrow X_{0F} = -\frac{L_x}{2} + \frac{\sqrt{\hbar eB}u_F}{eE(1-\beta^2)^{1/4}}\left(\sqrt{8}-\sqrt{6}\right)$$

(3.30)

Now, we determine the starting ($l_0$) and final ($l_F$) point of $l$ for this case:

$$l_0 = -\frac{\Phi}{2\Phi_0} + \frac{L_y}{2\pi l_B}\frac{\sqrt{6}\beta}{(1-\beta^2)^{1/4}} \quad (3.31)$$

$$X_0 = l_B^2 \frac{2\pi l_F}{L_y} - \frac{\sqrt{6}l_B\beta}{(1-\beta^2)^{1/4}}$$

$$\Rightarrow l_F = -\frac{\Phi}{2\Phi_0} + \frac{L_y}{h}\frac{B\sqrt{\hbar eB}u_F}{E(1-\beta^2)^{1/4}}\left(\sqrt{8}-\sqrt{6}\right) + \frac{L_y E}{hu_F\sqrt{eB}}\frac{e\sqrt{6\hbar}}{(1-\beta^2)^{1/4}}$$

(3.32)

The number of states in L.L. $n=3$ is given by:

$$l_F - l_0 = \frac{L_y}{h}\frac{B\sqrt{\hbar eB}u_F}{E(1-\beta^2)^{1/4}}\left(\sqrt{8}-\sqrt{6}\right), \quad (3.33)$$

and the total number of states under Fermi energy is:

$$g = \underbrace{\frac{\Phi}{\Phi_0}}_{n=0} + 2\frac{\Phi}{\Phi_0} + \frac{L_y}{h}\frac{B\sqrt{\hbar eB}u_F}{E(1-\beta^2)^{1/4}}\left(\sqrt{8}-\sqrt{6}\right) \quad (3.34)$$

If all states are filled with electrons, then the electron number is (considering that we have four electrons in each case, except for *n*=0, in which we place two electrons in each state):

$$N = 10\frac{\Phi}{\Phi_0} + 4\frac{L_y}{h}\frac{B\sqrt{\hbar eB}u_F}{E(1-\beta^2)^{1/4}}\left(\sqrt{8}-\sqrt{6}\right) \quad \text{(3.35)}$$

This relation modifies Hall conductivity as:

$$\sigma_H = \frac{n_A e}{B} = N\frac{e}{SB} = 10\frac{e^2}{h} + 4\frac{e}{hL_x}\frac{\sqrt{\hbar eB}u_F}{E(1-\beta^2)^{1/4}}\left(\sqrt{8}-\sqrt{6}\right) \quad \text{(3.36)}$$

Now, by directly using eq. (3.9) and substituting in (3.36):

$$\frac{\sqrt{2e\hbar B}u_F}{(1-\beta^2)^{1/4}}z = eEL_x \Rightarrow E = \frac{\sqrt{2e\hbar B}u_F}{eL_x(1-\beta^2)^{1/4}}z \quad \text{(3.37)}$$

we conclude to:

$$\sigma_H = \frac{n_A e}{B} = N\frac{e}{SB} = 10\frac{e^2}{h} + 4\frac{e^2}{zh}\left(\sqrt{4}-\sqrt{3}\right) \quad \text{(3.38)}$$

demonstrating the possibility of irrational quantization of the Hall conductance (always if the *E*-field is sufficiently strong).

## 4. Considering the QHE-breakdown in high injected currents

At this point one may wonder about the relevance of the above results, as these do not seem to take into account mechanisms associated with the breakdown of QHE under non-equilibrium conditions, i.e. due to high injected current densities [14-16]. Indeed, in the case of hundreds of μA (0.6-0.9 mA, taken from ref. [16], the exact value depending on the filling factor) flowing through the sample in the *y*-direction, the longitudinal resistivity is different from zero by orders of magnitude, resulting in an inevitable dissipation of energy, and the Hall conductivity deviates from its exact quantized values.

One of the reasons behind this deviation is the thermal instability that happens due to large current densities flowing continuously for a period of time through the material. The amount of heat gained by the electrons per unit time per unit area due to the electric field is given by

$$Q_{in} = \sigma_{yy}E^2$$

This amount of heat is then passed to the lattice structure, ending up in the Helium coolant. This means that there is a gradient of temperature between the electrons and the atoms forming the crystal that relates the heat transferred to the atoms through the following equation:

$$Q_{out} = k(T_e - T_h)$$

When the electric field exceeds a critical value, the amount of heat gained by the electrons ($Q_{in}$) is always greater than the amount of heat transferred to the atoms ($Q_{out}$) and the excess in thermal energy results in an energy excitation in the order of the cyclotronic energy, destroying the stability of the quantum Hall conductivity plateaus.

It should be noted that the present work takes into account the applied *E*-field directly within the energy spectrum, and there is no need to investigate the heating of the electron gas separately. It is contained a priori in the thermodynamic formulation that we have used to derive our results. Furthermore, the results shown in this paper can be relevant at even relatively low *E*-fields, or low injected current densities as well. To see this, we may assign some numbers to our mathematical relations, starting from eq. (2.1) that gives the limit of strong *E*-field: $E > \hbar\omega/eL_y$ for *z*>1. For a strong magnetic field of 40 Tesla, and a macroscopic length $L_y$ of 1 cm, we have:

$$E = \hbar\omega/eL_y = \hbar B/mL_y = 1.054 \times 40/91 = 0.46 V/m.$$

Above this critical *E*-field (and therefore for low Hall *E*-field) all the inter-L.L. overlaps that we considered can occur and the need for introducing breakdown effects is therefore eliminated. In the non-equilibrium breakdown regime of the QHE state in Graphene, it has been verified that not all L.Ls experience the breakdown [17]. For example, the *n=0* L.L. remains robust under the influence of a relatively strong *E*-field. Therefore, the breakdown in Graphene is a filling factor-dependent phenomenon, with different critical electric fields for each filling factor in a way that we may always impose a low enough *E*-field to achieve the results derived in this paper. The stronger the *B*-field is, the weaker the *E*-field gets, much lower than the breakdown limit.

Apart from the above mentioned electron-heating mechanism we point out, for completeness, that percolation of incompressible regions [18] has also been invoked for the QHE-breakdown, as well as the possible existence of compressible regions in the bulk [19], which however do not seem to offer insights relevant to the present calculation.

Finally, there exists a small number of publications in the literature that seem to be close to our exact solution (although they deviate from it); they also find fractional quantization of the Hall conductivity by seriously taking into account a conceptual difference between the externally applied *E*-field and the Hall-electric field (the one that is formed in the (classical) steady state, as opposed to the field being applied). They take into account the non-negligible effect of the Hall *E*-field on the electronic density of states, which is further broadened by the presence of this Hall *E*-field, see i.e. ref. [20]. Just like interactions lead to a splitting of L.Ls (as is well-known in the Fractional QHE), the Hall electric field succeeds in a similar manner to divide each Landau Level into many sub-levels without any introduction of impurities or interactions. Fractional quantization is then observed in higher L.Ls (due to the density of states-broadening), but not in the lowest L.L. (where interactions are dominant, and a Composite Fermion approach – hence a passage to the so-called $\Lambda$-Levels – must be used instead (see [9] for a practical use of $\Lambda$-Levels)). However, strong anisotropies have been observed at these fractions [17] which suggests that they need to be considered more carefully as functions of the Hall *E*-field.

By way of comparison with the above, we must state that in the present work we have avoided the notion of a Hall electric field, as this is a built-in consequence of the presence of the *B*-field in the (classical) steady state (or in a stationary state in the corresponding quantum problem), or, alternatively, as this can be viewed as a response of the system to an external voltage (or the externally applied *E*-field) which is already fully taken into account in our physical picture (Hamiltonian) and in our associated calculations. We use, instead, the "physical broadening" given by the presence of the external *E*-field in order to describe the fractional quantization. Furthermore, with respect to the possible robustness of our results (such as eq. (2.18) for a conventional system, that gives the Hall conductivity value for a certain magnetic field and fixed particle number), we recall that topological protection and exact quantization of $\sigma_H$ in a high *E*-field is not guaranteed for our system in the thermodynamic limit. This is better clarified at least in a visual manner in Figures 2.8a and 2.8b (where the number of edge states plays a role), but it further needs inclusion of other factors not taken into account here, like the impurity potential; this, however, is actually expected to have a positive effect – it is expected to eventually broaden the electronic density of states, and, at least, assist in establishing the topological stability of the system (in a similar manner as this occurs in the ordinary Integer-QHE).

## 5. Conclusions

We have shown that by finding the optimal energy at zero temperature in the case of conventional semiconductors, fractional Hall values appear at the points of jumps of the Fermi energy. This does not necessarily mean that plateaux appear. Inclusion of impurity potential may lead to quantized plateau structure even in the case of strong *E* field, due to further broadening of L.Ls, isolating extended from localized states. We should point out that it is not the gap closing that causes plateau disappearance, but rather the fact that we have ignored the impurity potential in our calculations. We have also calculated analytically, using ground state energy considerations the total internal energy, magnetization and polarization as functions of the electromagnetic field. The associated de Haas-van Alphen oscillation periods are also influenced by the presence of the electric field in a specific quantitative manner. A corresponding exact calculation in a pseudo-relativistic system, such as Graphene, is more involved but it has also been carried out here in detail. An immediate result of our toy model is the possibility of irrational Hall values, although further investigation is required (i.e. inclusion of disorder) in order to see if these effects survive under realistic conditions.